\documentstyle[multicol,prl,aps,psfig,tighten]{revtex}
\newcommand{\BEQ}{\begin{equation}}     
\newcommand{\BEA}{\begin{eqnarray}}
\newcommand{\EEQ}{\end{equation}}       
\newcommand{\EEA}{\end{eqnarray}}
\newcommand{\D}{{\rm d}}

\renewcommand{\vec}[1]{{\bf{#1}}}       
\def\vecr{{\bf r}}
\def\qea{q_{\rm EA}}

\begin{document}

\input epsf.sty

\draft

\widetext

\title{Local scale invariance as dynamical space-time symmetry in 
phase-ordering kinetics}

\author{Malte Henkel$^a$ and Michel Pleimling$^{b}$}

\address{
$^a$ Laboratoire de Physique des Mat\'eriaux,$^{*}$
Universit\'e Henri Poincar\'e Nancy I, B.P. 239,\\
F -- 54506 Vand{\oe}uvre l\`es Nancy Cedex, France\\
$^b$ Institut f\"ur Theoretische Physik I, Universit\"at Erlangen-N\"urnberg,
D -- 91058 Erlangen, Germany}
\maketitle

\begin{abstract}
The scaling of the spatio-temporal response of coarsening systems is
studied through simulations of the $2D$ and $3D$ Ising model with Glauber 
dynamics. The scaling functions agree with the prediction of local
scale invariance, extending dynamical scaling to a space-time dynamical
symmetry.  
\end{abstract}
\pacs{05.20-y, 64.60.Ht, 75.40.Gb, 11.25.Hf}

\begin{multicols}{2}
\narrowtext
Ageing phenomena are observed in a broad variety of systems
with slow relaxation dynamics. Useful insight may often
be gained via the consideration of simple ferromagnetic models (rather than
genuinely glassy systems), quenched to a temperature $T<T_c$ from 
a disordered initial state into the low-temperature phase. For their study 
two-time quantities, rather than the more usual one-time quantities, 
are particularly useful, see \cite{Bou00,God02,Cug02} for recent reviews. 
Those display dynamical self-similarity in the ageing regime where the 
order-parameter autocorrelator $C(t,s)=\langle\phi(t)\phi(s)\rangle$ 
decays from its plateau value $\qea=M_{\rm eq}^2$ to zero ($M_{\rm eq}$ is 
the spontaneous magnetization). In this regime both~$s$ and $\tau=t-s$ are much 
larger than the microscopic time scale (set to unity). Similarly, 
the autoresponse function, defined as 
$R(t,s)=\delta\langle\phi(t)\rangle/\delta h(s)$ with $t>s$ 
shows a scaling behaviour such that the scaling laws  
\BEQ
C(t,s)\simeq M_{\rm eq}^2 f_C(t/s) \;\; , \;\; R(t,s)\simeq s^{-1-a} f_R(t/s)
\label{gl:CR}
\EEQ
with $f_C(1)=1$ are found to hold for a broad range of 
models \cite{Bou00,God02,Cug02}. For $x\gg1$, the scaling functions usually 
fall off as $f_C(x)\sim x^{-\lambda_C/z}$ and $f_R(x)\sim x^{-\lambda_R/z}$,
where $z$ is the dynamic critical exponent, $\lambda_C$ is the 
autocorrelation exponent \cite{Fis88,Hus89} and $\lambda_R$ is the
autoresponse exponent \cite{Pic02}. 
We shall focus here on phase-ordering in the Glauber-Ising model 
in $d>1$ dimensions where $a=1/z=1/2$ is expected \cite{Ber99,God02}. 
Recent arguments \cite{Cor01} 
leading to $a=1/4$ in the $2D$ Glauber-Ising model have been rejected through
a detailed study of the scaling of $R(t,s)$, the results of which reconfirm 
$a=1/2$ \cite{Hen02a}.

Eq.~(\ref{gl:CR}) states that the 
two-time quantities evaluated at the {\em same} spatial location 
transform covariantly under a global rescaling of time $t\to b t$ with
$b$ constant. Recently, it has been proposed that the response functions
should transform covariantly under {\em local} scale transformations with 
$b=b(t)$, but with time-translations excluded \cite{Hen02}. By analogy with
conformal invariance, in particular covariance of $R$ under the so-called 
`special' transformations which transform time as $t\to t/(1+\gamma t)$,
is assumed. If that is the case, the exact autoresponse scaling function 
becomes \cite{Hen02,Hen01} 
\BEQ \label{R}
f_R(x) = r_0 x^{1+a-\lambda_R/z} (x-1)^{-1-a}
\EEQ
and depends on the universality classes
only through the values of the exponents $a$ and $\lambda_R/z$, while 
$r_0$ merely is a normalization constant. This prediction of local
scale invariance (LSI) has been confirmed in a variety of physically
very different models, see \cite{Hen02} and references therein. Still, better 
evidence in favour of LSI than mere phenomenology would be desirable. 

The origin of LSI can be understood in the special case $z=2$ 
(which is realized in all cases of phase-ordering kinetics we are
going to consider \cite{Notiz1}). A Ward identity can be written down 
such that if the system is known to be (i) scale-invariant with $z=2$ and 
(ii) Galilei-invariant, then its invariance under the full group of local 
scale transformations follows \cite{Hen03}. Therefore, scale invariance and
Galilei invariance appear as the building blocks for LSI. So far, no 
{\it direct} test of Galilei invariance in phase-ordering systems was 
ever performed.

Direct tests of Galilei invariance require the 
study of the time- {\em and} space-dependence of response functions. For
phase-ordering, it is natural to expect a scaling behaviour of the linear
response function
\begin{displaymath}
R(t,s;\vec{r}) = \left.\frac{\delta\langle\phi_{\vecr}(t)\rangle}
{\delta h_{{\vec 0}}(s)}\right|_{h=0} = s^{-1-a} F_R\left(\frac{t}{s},
\frac{\vec{r}}{(t-s)^{1/z}}\right)
\end{displaymath}
with the scaling function $F_R(x,\vec{u})$ and $F_R(x,\vec{0})=f_R(x)$. For
$z=2$, Galilei invariance (combined with the Ward identity of LSI) 
predicts \cite{Hen94,Hen02,Hen03}
\BEQ \label{RR}
R(t,s;\vec{r}) = R(t,s) \exp\left(-\frac{\cal M}{2} 
\frac{\vec{r}^2}{t-s}\right)
\EEQ
where $R(t,s)$ is the autoresponse function given by eqs.~(\ref{gl:CR},\ref{R})
and ${\cal M}$ is a direction-dependent non-universal constant. 
Eq.~(\ref{RR}) gives the full spatio-temporal scaling of the linear response. 
We shall sucessfully test it in the $2D$ and $3D$ kinetic Glauber-Ising models 
and provide thereby the first direct evidence of Galilei invariance in a 
phase-ordering system. Since the $2D$/$3D$ Glauber-Ising model cannot be 
reduced to a free-field theory, it is highly non-trivial that its {\em exact} 
response function takes the simple free-field form eq.~(\ref{RR}).

As precise simulational data for the autoresponse $R(t,s)$ are difficult
to obtain, it is convenient to study instead the thermoremanent 
magnetization \cite{Bar98} 
\BEQ \label{Mtrm} 
T M_{\rm TRM}(t,s)/h_{(0)} = \rho(t,s) =T  \int_{0}^{s} \!\D u\, R(t,u) 
\EEQ 
We consider the Ising model on a hypercubic lattice, with periodic boundary
conditions and the Hamiltonian ${\cal H}=-\sum_{(\vec{i},\vec{j})} 
\sigma_{\vec{i}}\sigma_{\vec{j}}$. We use heat-bath dynamics defined through
the stochastic rule
\BEQ \label{heat}
\!\!\sigma_{\vec{i}}(t+1) = \pm 1 \mbox{\rm with probability $\frac{1}{2}\left[
1\pm \tanh(h_{\vec{i}}(t)/T)\right]$}
\EEQ
with the local field 
$h_{\vec{i}}(t)=\sum_{\vec{a}(\vec{i})}\sigma_{\vec{a}}(t)$
and where $\vec{a}(\vec{i})$ runs over the nearest neighbours of the 
site $\vec{i}$. $M_{\rm TRM}(t,s)$ is measured by applying a quenched 
spatially random magnetic field $\pm h_{(0)}$ for times between the quench at
$t=0$ and the waiting time $s$ \cite{Bar98}. The presence of this external 
field then changes the local field in eq.~(\ref{heat}) to 
$h_{\vec{i}}(t)=\sum_{\vec{a}(\vec{i})}\sigma_{\vec{a}}(t) \pm h_{(0)}$. 
 
It has been understood recently that the scaling behaviour of $M_{\rm TRM}$ 
for $s\gg 1$ is not a simple power law but rather shows a cross-over 
behaviour \cite{Zip00,Cor01}. It is of the form \cite{Hen02a} 
\BEQ \label{gl:rho}
\rho(t,s) = r_0 s^{-a} f_M(t/s) + r_1 s^{-\lambda_R/z} g_M(t/s)
\EEQ
provided the system was initially prepared at infinite temperature. 
Here the correction term can be estimated as $g_M(x) \simeq x^{-\lambda_R/z}$,
while the scaling function $f_M(x)$ can be found from local scale
invariance using eqs.~(\ref{gl:CR},\ref{R},\ref{Mtrm}) 
and is given explicitly by eq.~(5.47) in \cite{Hen02}. 
Finally, $r_{0,1}$ are normalization constants. In practice, it
turns out that $\lambda_R$ is quite close to its lower bound $d/2$ and 
in particular in $2D$, the correction to the leading scaling behaviour is
sizeable. Before any meaningful study of the spatio-temporal response can be 
carried out, the correction term must be subtracted off.  

As explained in \cite{Hen02a}, this can be carried out by fixing a value of
$x=t/s$. Then $r_{0,1}$ are obtained by fitting eq.~(\ref{gl:rho})
to the computed thermoremanent magnetization $M_{\rm TRM}$. 
In order to illustrate the 
quality of this procedure, we show in figure~\ref{Abb1} data for the 
scaling function $f_M(x)$ obtained after subtraction of the correction term
$r_1 s^{-\lambda_R/z} g_M(t/s)$ from the integrated autoresponse $\rho$
of the $2D$ and $3D$ Glauber-Ising model. Statistical error bars in $2D$ are 
smaller than the symbol size and in $3D$ are of the order of the scatter in the
data. In table~\ref{tab1}, we list the
values of $\lambda_R$ \cite{Fis88,Hum91} and the constants $r_{0,1}$ 
which were determined at $x=7$. We find a nice data collapse and a clear 
agreement with the LSI prediction.

Having thus checked the expected scaling of the autoresponse function, we can
now turn towards the spatio-temporal response. Again, we consider the
integrated response rather than $R(t,s;\vec{r})$ because it is considerably 
less affected by noise. In order to fix the non-universal parameter $\cal M$
in (\ref{RR}), we form
\BEQ
\rho_0(t,s;\vec{r}) = T \int_0^{s} \!\D\tau\, R(t,s-\tau;\vec{r})
\EEQ
which is measured on the lattice by keeping a small random magnetic field 
$\pm h_{(0)}$ until the waiting time $s$ at the site $\vec{0}$ and then 
observing the thermoremanent magnetization 
$M_{\rm TRM}(t,s;\vec{r})=h_{(0)}\rho_0(t,s;\vec{r})/T=
\overline{\langle h_{\vec{0}}\,\sigma_{\vec{r}}(t)\rangle}/T$ 
at a different site $\vec{r}$ (where $\langle \cdots \rangle$ indicates 
an average over the thermal noise whereas the bar means an average over 
the random field distribution). 
We computed the spatially and temporally integrated response for the
Ising model with Glauber dynamics in two and three dimensions. The 
two-dimensional systems usually contained $300^2$ spins on a square
lattice, whereas in three dimensions cubic lattices with typically $60^3$
sites were considered. Some Monte Carlo simulations were also 
done for other system sizes in order to check against finite-size effects.
Since $\rho_0(t,s;\vec{r})$ is very noisy, one must 
average over a large number of runs with different realizations of
the thermal noise and of the spatially random magnetic field. 
For every waiting time we avaraged over at least 50000 different runs.
The whole study took about $2\cdot10^5$ CPU hours on a SGI Origin 3800
parallel computer.

Following the same lines as in \cite{Hen02a}, we arrive at the 
scaling form\cite{Note:qe}, where $\vec{r}$ varies along a fixed direction
on the lattice
\BEA 
\rho_0(t,s;\vec{r}) &\simeq& \rho_{\infty}(t-s;\vec{r}) 
+ r_0 s^{-a} F_0\left(\frac{t}{s}, {\cal M}\frac{r^2}{s}\right) 
\nonumber \\
& & + r_1 s^{-\lambda_R/z} G_0\left(\frac{t}{s}, {\cal M}\frac{r^2}{s}\right) 
\label{rho0}
\EEA 
where the universal scaling functions read 
\BEA 
F_0(x,y) &=& \int_0^1 \!\D v\, 
\exp[ -y/2(x-1+v)]\, h_R(x,v)
\nonumber \\
G_0(x,y) &\simeq& x^{-\lambda_R/z} e^{-y/2x}
\EEA
with $h_R(x,v):=f_R\left(\frac{x}{1-v}\right) (1-v)^{-1-a}$ 
where $\cal M$ is a direction-dependent new parameter. In this work we
choose $T$ such as to keep anisotropy effects small and also to avoid the
cross-over to critical dynamics at $T=T_c$. This is illustrated in 
figure~\ref{Abb2}, for the $2D$ case at $T=1.5$ where we plot $F_0$ 
as function of $r^2/s$. 
For the displayed data points $r$ varies between 1 and $\sqrt{s}$
when going along the (10) direction and between $\sqrt{2}$ and
$\sqrt{2s}$ when going along the (11) direction.
At the considered temperature any anisotropies are very small and
the values of ${\cal M}$, determined separately in the two directions, 
coincide, ${\cal M}_{(10)}={\cal M}_{(11)}={\cal M}$. 
Similar results also hold in $3D$ at $T=3$.
We fix ${\cal M}$ at $x=7$ (see table~\ref{tab1} and inset) using (\ref{rho0}) 
and then obtain a parameter-free prediction for other values of $x$. 
This is shown in the inset of figure~\ref{Abb2}, where the 
LSI-prediction is compared to numerical 
data for two additional values of $x$ and two different 
waiting times. Since by now all non-universal parameters are fixed,
this constitutes by itself a quantitative test of local scale invariance.

A fuller test of the spatio-temporal response is obtained by considering
the spatially and temporally integrated response function
\BEQ \label{drdO}
\frac{\D \rho(t,s;\mu)}{\D \Omega} = 
T \int_0^s \!\D u\, \int_0^{\sqrt{\mu s}} \!\D r\, r^{d-1} R(t,u;\vec{r})
\EEQ 
where the space integral is along a straight line of length 
$\Lambda=\sqrt{\mu s}$ but we do allow for the possibility of an
anisotropy as a function of the solid angle $\Omega$. Such anisotropies are
known to occur if $T<T_c$ \cite{Lee99}. 
As before, we derive from (\ref{RR}) the
scaling form\cite{Note:qe}
\BEA
\lefteqn{
\frac{\D \rho(t,s;\mu)}{\D \Omega} = \rho^{(1)}(t-s,\vec{r}^2) 
} \label{rho}
\\ &+& 
r_0 s^{d/2-a} \rho^{(2)}(t/s,\mu) +
r_1 s^{d/2-\lambda_R/z} \rho^{(3)}(t/s,\mu)
\nonumber 
\EEA
with the explicit scaling functions
\BEA
\rho^{(2)}(x,\mu) &=& \frac{\mu^{d/2}}{d} \int_0^1 \!\D v\, h_R(x,v) 
{\cal F}_d\left(\frac{{\cal M}\mu}{x-1+v}\right) 
\nonumber \\
\rho^{(3)}(x,\mu) &\simeq& \frac{\mu^{d/2}}{d}\, x^{-\lambda_R/z}
{\cal F}_d\left(\frac{{\cal M}\mu}{x}\right)
\label{rho2}
\\
{\cal F}_d(y) &=& e^{-y/2}\, {_1\mbox{\rm F}_1}\left(1,1+\frac{d}{2};
\frac{y}{2}\right)
\nonumber 
\EEA
This is the general expression for the scaling of the spatio-temporally
integrated response function. 
If we fix $\mu$ and let $x=t/s$ vary, the form of the scaling function of
$s^{a-d/2} \D\rho/\D\Omega$ merely depends on $\mu$. We stress that the 
exponent $\lambda_R$ and the free parameters $r_{0}, r_{1},{\cal M}$ 
are now all fixed such that there remains no free fitting parameter at 
all when comparing (\ref{rho}) with our numerical data. 

It is of interest to compare the maximal distance $r_{\rm max}$ 
accessed by our simulation
with the physical length scale of the problem, for example the domain size
$L(s)$ at the instant the magnetic field is turned off. From the correlation
function, in $2D$ we estimate $L(s)\approx 3,6,9.5$ for $s=25, 100, 225$,
respectively, with $h_{(0)}=0.05$. 
In $3D$, we find $L(s)\approx 2.1, 3.3, 4$ for $s=25,64,100$. 
As expected $L(s)\sim s^{1/z}$, with $z\approx 2$. 
The value $\mu=4$ ($\mu=2$) corresponds to $r_{\rm max}/L(s)\approx 3.2$ (3.4) in 
$2D$ ($3D$), respectively and we see that the data probe the large-distance 
region, well beyond the domain size $L(s)$.

The integrated spatio-temporal response in $2D$ at $T=1.5$ 
is shown in figure~\ref{Abb3} for two values of $\mu$ and where the correction
$\rho^{(3)}$ was already subtracted off. We find a nice
scaling behaviour over the whole range of waiting times we could consider.
We stress that the agreement between local scale invariance (\ref{RR}) and our
data for several values of $\mu$ is a real test, since no free parameter
remains. In particular, both the height and the position of the maximum
of the scaling function for $\mu=2$ and $4$ is completely fixed. 
We can conclude that LSI is fully
vindicated. This is the first time that the exact functional form of the
scaling function of the spatio-temporal response of a generic non-equilibrium 
spin system is found. 

Similarly, the integrated spatio-temporal response in $3D$ at $T=3$ is 
shown in figure~\ref{Abb4}. Again, a nice scaling behaviour in perfect
agreement with LSI is found. 

The prediction (\ref{RR}) is also verified in a few exactly solvable models
with a non-conserved order parameter and undergoing phase-ordering kinetics
with $z=2$. In particular, (\ref{RR}) holds in the $d$-dimensional kinetic
spherical model with $d>2$ quenched to either $T<T_c$ or $T=T_c$ and 
independently of initial correlations \cite{New90,Jan89,God00b,Zip00,Pic02}.

Summarizing, we presented the first quantitative study of the scaling of the
spatio-temporal response in coarsening systems. For the first time, 
we find direct evidence for Galilei invariance in 
the ageing regime of a phase-ordering system. This
is a new {\em dynamic symmetry} whose presence in a non-equilibrium phase
transition was not anticipated. Our result provides a strong indication that 
LSI is indeed a true {\em space-time} dynamical symmetry of statistical systems 
undergoing phase-ordering kinetics. 

We thank B.P. Vollmayr-Lee and C. Godr\`eche for discussions. 
This work was supported by the Bayerisch-Franz\"osisches Hochschulzentrum 
(BFHZ) and by CINES Montpellier (projet pmn2095).

\vspace{-5mm}

{
\begin{figure}[htb]
\centerline{\epsfxsize=3.25in\ \epsfbox{
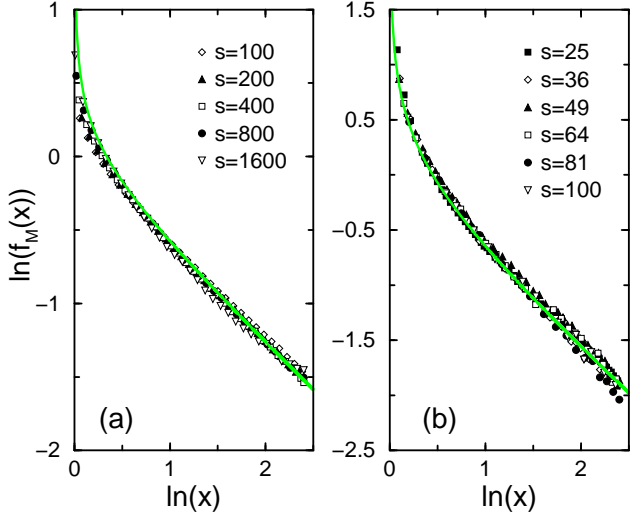}
}
\caption{Scaling of the integrated response function $\rho$ for the
Glauber-Ising model. (a) $2D$ at temperature $T=1.5$ (b) $3D$ at $T=3$.
The symbols correspond to different waiting times. The full curves are
obtained by integrating~{\protect(\ref{R})}.
\label{Abb1}}
\end{figure}
}

{
\begin{figure}[htb]
\centerline{\epsfxsize=3.25in\ \epsfbox{
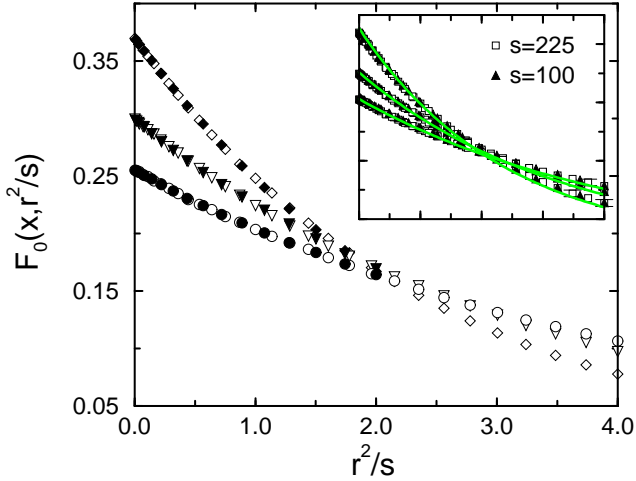}
}
\caption{Scaling of the temporally integrated response function $\rho_0$
for the $2D$ Glauber-Ising model at $T=1.5$ and $s=225$. 
Data obtained along the (10)
resp. (11) direction are shown by open resp. filled symbols. 
Diamonds: $x=5$, triangles: $x=7$, circles: $x=9$. Inset: Determination 
of the mass $\cal M$, see main text. Some typical error bars are shown
in the inset.\label{Abb2}}
\end{figure}
}

{
\begin{figure}[htb]
\centerline{\epsfxsize=3.25in\ \epsfbox{
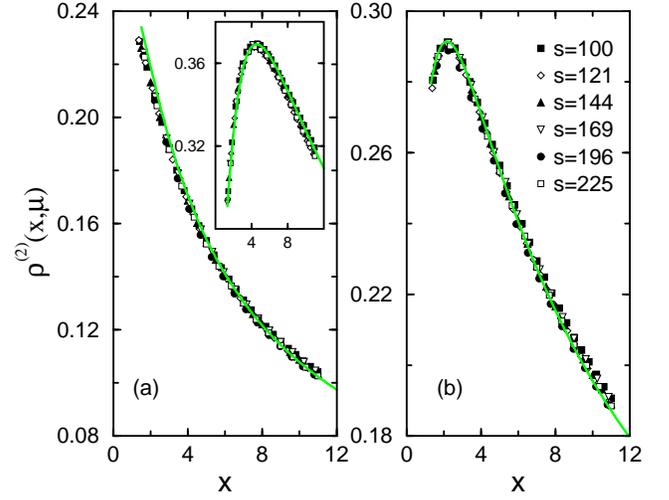}
}
\caption{Scaling of the integrated spatio-temporal response function 
$\rho$ for the $2D$ Glauber-Ising model at $T=1.5$
and at (a) $\mu=1$ (inset: $\mu=4$) and (b) $\mu=2$
(integrating (a) along the (10) direction and (b) along the (11) direction). 
The full curves are from eq.~{\protect(\ref{rho2})}. Error bars are smaller 
than the symbol size.\label{Abb3}}
\end{figure}
}

{
\begin{figure}[htb]
\centerline{\epsfxsize=3.25in\ \epsfbox{
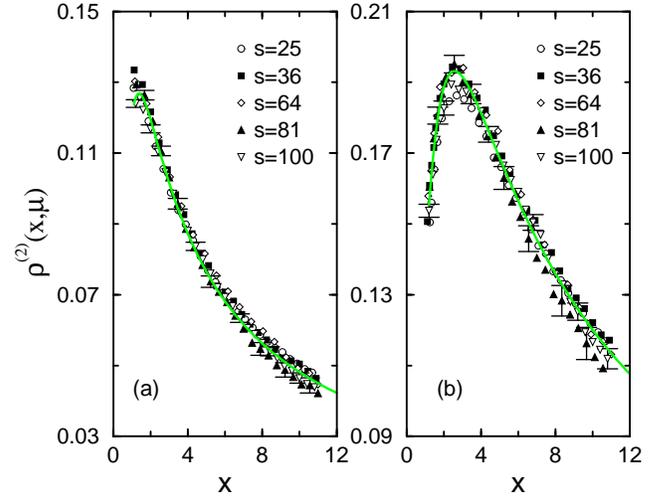}
}
\caption{Scaling of the integrated spatio-temporal response function 
$\rho$ for the $3D$ Glauber-Ising model at $T=3$
and at (a) $\mu=1$ and (b) $\mu=2$
(integrating (a) along the (100) direction and (b) along the (110) direction). 
The full curves are from eq.~{\protect(\ref{rho2})}. Some typical error bars 
are also shown.\label{Abb4}}
\end{figure}
}

\begin{table}[ht]
\caption{Values of the autoresponse exponent $\lambda_R$ 
and of the parameters $r_0$, $r_1$ and ${\cal M}$ in the 
Glauber-Ising model in two dimensions at $T=1.5$ and in 
three dimensions at $T=3$. 
\label{tab1}}
\begin{tabular}{|l|cccc|} 
     & $\lambda_R$ & $r_0$    & $r_1$            & ${\cal M}$     \\ \hline
$2D$ & 1.26 & $2.65 \pm 0.05$ & $-2.76\pm 0.05$  & $4.08 \pm 0.04$\\ 
$3D$ & 1.60 & $0.31 \pm 0.02$ & $~~0.61\pm 0.02$ & $4.22 \pm 0.05$\\ 
\end{tabular}
\end{table}

\end{multicols}


\begin{references}
\bibitem{Notiz1} For $z=2$ the LSI-group is the Schr\"odinger group, 
the max. kinematic group of the free Schr\"odinger equation.
\end{references}
\end{document}